\documentclass[9pt, nofootinbib, notitlepage, twocolumn]{revtex4-1}

\usepackage{graphicx} 
\usepackage[utf8]{inputenc}

\usepackage{fontenc}

\usepackage{amsmath}
\usepackage{amssymb}
\usepackage{amsthm}
\usepackage{hyperref}
\usepackage{amsmath,amssymb,amsthm, graphicx, array,epsfig, exscale, dsfont, amsfonts, verbatim, fancyhdr, natbib, bbm,mathrsfs,textcomp,listings, epsfig, mathtools}

\usepackage{color}

\usepackage{slashbox}

\newcommand{\one}{\mbox{$1 \hspace{-1.0mm}  {\bf l}$}}

\newcommand{\ket}[1]{\left\vert{#1}\right\rangle}

\newcommand{\avg}[1]{\left<#1\right>}

\date{\today}

\begin{document}

\title[Finite-range multiplexing enhances quantum key distribution via quantum repeaters]{Finite-range multiplexing enhances quantum key distribution via quantum repeaters}
\author{Silvestre Abruzzo, Hermann Kampermann, Dagmar Bru{\ss}}
\address{Institute for Theoretical Physics III, Heinrich-Heine-Universit\"at D\"usseldorf, Universitätsstr. 1, 40225 D\"usseldorf, Germany}

\begin{abstract}
Quantum repeaters represent one possible way to achieve long-distance quantum key distribution. Collins et al. in [Phys. Rev. Lett. 98, 060502 (2007)] proposed multiplexing as method to increase the repeater rate and to decrease the requirement in memory coherence time. Motivated by the experimental fact that long-range connections are practically demanding, in this paper we extend the original quantum repeater multiplexing protocol to the case of short-range connection. We derive analytical formulas for the repeater rate and we show that for short connection lengths it is possible to have most of the benefits of a full-range multiplexing protocol. Then we incorporate decoherence of quantum memories and we study the optimal matching for the Bell-state measurement protocol permitting to minimize memory requirements. Finally, we calculate the secret key rate and we show that the improvement via finite-range multiplexing is of the same order of magnitude as via full-range multiplexing.  
\end{abstract}

\maketitle

\section{Introduction}

Quantum key distribution (QKD) \cite{bennett1984quantum, ekert1991quantum, Scarani:2009} allows two parties to share a secret key which might be used for applications in cryptography. The preferred quantum systems used for transmitting information are photons. These can be generated, distributed and measured fairly easily with standard technology. However, photons are usually transmitted through optical fibers and due to absorption the maximal distance where QKD is feasible is around 150 km \cite {Scarani:2009}. In order to overcome this problem the concept of quantum repeaters can be used \cite{briegel_quantum_1998, duer_quantum_1999}.  For increasing  the final repeater rate and the final fidelity many variations of the original protocol have been investigated \cite{duan_long-distance_2001, van_loock_hybrid_2006, sangouard_quantum_2011}, where one of the influential generalizations is  \emph{multiplexing} \cite{Collins2007}.

\begin{figure}
 \includegraphics[width=.5\textwidth]{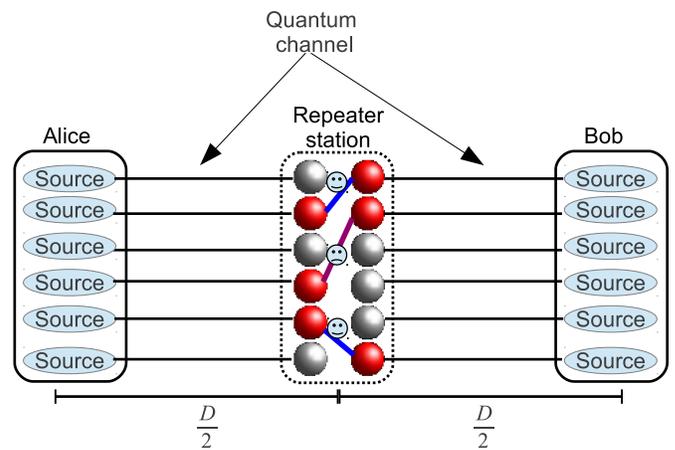}
\caption{(Color online) Alice and Bob are equipped with single-photon sources. Each source is connected through an optical fiber to a quantum memory in the repeater station. Red spheres represent filled quantum memories whereas gray spheres represent empty quantum memories.  In this example the maximal connection length is one, therefore the connections indicated in blue are allowed and the magenta one is forbidden.}
\label{fig:MultiplexingScheme} 
\end{figure}

In fig.~\ref{fig:MultiplexingScheme} we show a typical set-up of a quantum repeater with multiplexing. Alice and Bob have many single-photon sources which are connected to a quantum memory in the repeater station. Optical fibers are lossy, therefore after one attempt some quantum memories are filled up (red spheres in the picture) and some are empty (gray spheres).
One possibility is to perform Bell-state measurements (BSMs) only between parallel quantum memories; the second possibility, which is called multiplexing is to allow BSMs between two arbitrary quantum memories of the two arrays. In Ref.~\cite{Collins2007} the authors give an analytical formula for the entanglement production rate with multiplexing when quantum repeaters with two segments are considered. The conclusion of \cite{Collins2007} was that multiplexing gives only a modest improvement on the rate w.r.t. the case of parallel connections. However, it improves significantly the requirements on memory decoherence. In Ref.~\cite{jiang_fast_2007} a new protocol based on the Duan-Lukin-Cirac-Zoller protocol \cite{duan_long-distance_2001} has been studied and it has been found numerically that $R_{M}\approx R_{P}^{1.12}$ where $R_{M}$ is the rate using multiplexing and $R_{P}$ is the 
rate using parallel connections. Other works concerning multiplexing include Ref.~\cite{razavi_quantum_2009} which studied the repeater rate and the final fidelity in the limit of large number of quantum memories, Ref.~\cite{NadjaHybrid} which derived an analytical formula for the average number of attempts necessary for performing the first connection and Ref.~\cite{munro2010quantum} where a new protocol based on multiplexing has been proposed.
\newcommand{\frmp}{FIRMP}
\newcommand{\crmp}{FURMP}

In this paper we assume a set-up with one repeater station, i.e. two segments (see fig.~\ref{fig:MultiplexingScheme}).  We consider multiplexing when few quantum memories are used.  We introduce the \emph{finite-range multiplexing protocol} (\frmp), which is motivated by the fact that long-range connections are experimentally demanding \cite{leibrandt2009demonstration, sangouard2009quantum, kumph2011two}.  We provide analytical formulas for the repeater rate using the \emph{full-range multiplexing protocol} (\crmp) and the \frmp. Then we investigate  quantum memory decoherence, and we study numerically the optimal algorithm such that the memory requirements for QKD are minimized.

The manuscript is organized as follows. In sec.~\ref{sec:protocol} we introduce quantum key distribution and the quantum repeater protocol with  finite-range multiplexing. Moreover, we describe different Bell-state measurement strategies. In sec.~\ref{sec:reprate} we derive analytical formulas for the repeater rate in the case of deterministic and probabilistic Bell-state measurements. In sec.~\ref{sec:decoherence} we show how to minimize the memory requirement such that quantum key distribution is still possible and in sec.~\ref{sec:scr} we calculate the secret key rate. Finally, in sec.~\ref{sec:conclusions} we summarize the results and outline possible future developments.

\section{The protocol\label{sec:protocol}}
\subsection{General description}
Alice and Bob are two parties at a distance $D$ who want to create a secret key using QKD. Throughout the present paper we consider that they use a quantum repeater with two segments, i.e. one repeater station. This set-up is particularly important  because Alice and Bob do not necessarily need entanglement sources or quantum memories. Instead, single-photon sources or weak coherent pulse sources  are sufficient. This set-up resembles the measurement-device independent QKD protocol proposed in  \cite{PhysRevLett.108.130503, PhysRevLett.108.130502}. This protocol has been extended to the quantum repeater scenario with quantum memories in \cite{2013arXiv1306.3095A}. In this paper, the sources are supposed to be single-photon sources. However, the analysis for weak coherent pulse sources can be done following the methods developed in \cite{PhysRevLett.108.130503,2013arXiv1306.3095A}. We assume that the repeater station contains two arrays of $m$ quantum memories, where one side receives the photons sent from Alice and the other one 
receives the photons sent from Bob (see fig.~\ref{fig:MultiplexingScheme}).

\begin{figure}
 \includegraphics[width=.5\textwidth]{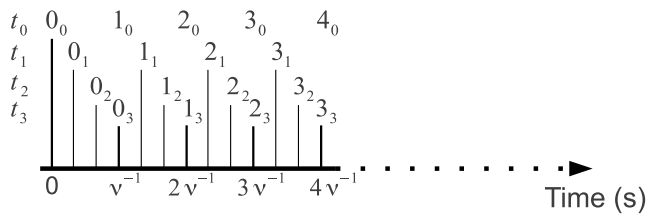}
\caption{(Color online) Description of our notation of time. Refer to the main text for the meaning of $t_i$. Bold vertical lines which are proportional to $\nu^{-1}$ represent the instant where  sources produce new photons. The quantity $\nu$ is the frequency of the source measured in pairs per second.}
\label{fig:TimeScheme} 
\end{figure}

In the following we give the steps of the multiplexing protocol with finite-range connection. We define   time variables denoted by  $t_{i}$ with integers $t\in[1, \infty)$ and $i=0,1,2,3$ interpreted as follows. The value of $t$ denotes the attempts of the sources to produce photons. As shown in fig.~\ref{fig:TimeScheme} the variable $t_i$ can be always related to the elapsed time in seconds from the beginning of the experiment by using the repetition frequency $\nu$ of the source which is measured in pairs per second. Therefore $1_0\equiv\nu^{-1}$, $2_0\equiv2\nu^{-1}$, etc. We will call the interval between $t_0$ and $(t+1)_0$ a time-bin. The subindex $i$ permits to describe instants contained in a  time-bin. At time $t_0=0_0$ all quantum memories are empty and 
the protocol is just starting. 
\begin{figure}
 \includegraphics[width=.5\textwidth]{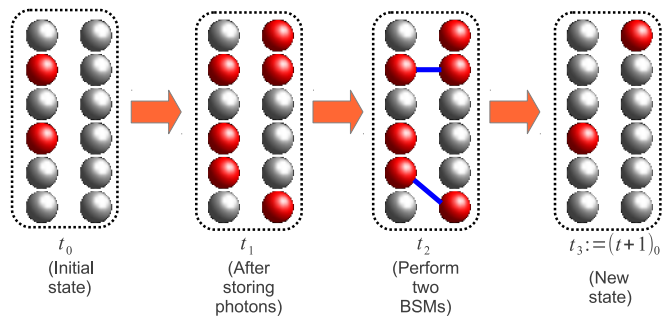}
\caption{(Color online) Representation of the repeater station, performing \frmp: Steps performed in the repeater protocol in one time-bin. A red (gray) sphere indicates that the quantum memory is filled (empty). We have $m=6$ quantum memories for each array and the connection length is $w=1$. In $t_2$ it is not possible to perform all possible connections due to the limited connection length. In the case of { \crmp } in $t_2$ three BSMs would be possible and in $t_3$ all quantum memories would be empty.}
\label{fig:scheme} 
\end{figure}
The steps are the following (see fig.~\ref{fig:scheme}):
\begin{enumerate}
 \item[at $t_0$]: Alice and Bob prepare randomly and independently $m$ random states each according to a chosen QKD protocol\footnote{For example, in the case of BB84 they prepare one of the four qubit states $\ket{\psi}\in\{\ket{0}, \ket{1}, \ket{+}, \ket{-}\}$ where $\ket{\pm}:=(\ket{0}\pm\ket{1})/\sqrt{2}$.}. They store the information regarding the preparation and they send the states to the repeater station. 
 \item[at $t_1$]: Arriving photons are stored and heralded in the corresponding quantum memory.  
 \item[at $t_2$]: The repeater station performs the maximal number  of Bell-state measurements $\ell$ compatible with the maximal connection length $w$. For  $w=0$ only BSMs in parallel are allowed and for $w=m-1$ any connection is possible. 
 \item[at $t_3$]: The measured quantum memories are again empty. Restart from $i=0$, i.e. $t_3\equiv(t+1)_0$.
\end{enumerate}

The repeater station communicates to Alice and Bob which pairs of quantum memories were used for the BSMs, as well as the measurements result. The protocol ends after a given number of rounds. After that,  Alice and Bob will execute the standard QKD protocol which consists of sifting, parameter estimation, error correction and privacy amplification \cite{Scarani:2009}. 

The advantage of a  multiplexing protocol occurs at time $t_2$ where BSMs are performed such that the number of connections is maximized. In the case of { \crmp } this corresponds to performing as many BSMs such that one array of quantum memories is completely empty. In the case of  { \frmp } the maximal number of connections can be found using the \emph{maximum cardinality bipartite matching}  \cite{west2001introduction,
kleinberg2006algorithm}, i.e. the maximum number of edges in a given bipartite graph such that each vertex has at most one neighbor. This problem can be explicitly formulated as follows: There is a list of filled quantum memories on the left. Each  quantum memory on the left may be connected to several quantum memories on the right, depending on the maximal connection length. This defines a bipartite graph. By solving the maximum cardinality bipartite matching algorithm we find the maximal number of possible connections. In the case of full-range multiplexing the optimal matching will always leave one of the two arrays completely empty. This is not the case with finite-range multiplexing. However, it remains true that at least one of two involved quantum memories has been filled in the same time-bin when the connection is done.  Obviously, there could be several possible matchings which maximize the number of connections. The chosen matching can have an influence on the final state fidelity due to  memory 
decoherence. Possible strategies for choosing a matching are discussed in the following. 

\subsection{Multiplexing strategies}

We can view the set-up in a repeater station as a bipartite weighted graph where the filled quantum memories are the vertices and the possible connections restricted by the maximal connection length are edges. To each vertex is assigned an integer value given by the arrival time of the stored photon. If the quantum memory is empty it does not represent a vertex.

\begin{figure}
 \includegraphics[width=.33\textwidth]{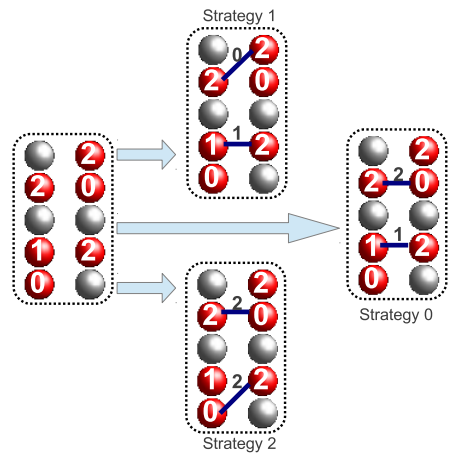}
\caption{(Color online) A red (gray) sphere indicates that the quantum memory is filled (empty). The number in the sphere represents the arrival time of the corresponding photon. On the left we have the situation at time $2_1$  (see fig.~\ref{fig:scheme}).  We consider $w=1$. On the right three possible  matching strategies are shown. Blue lines indicate the difference between the arrival times of the photons. It is possible to see the schemes on the right side as weighted bipartite graphs: red spheres are vertices and blue edges have indicated weights.}
\label{fig:decoherencescheme} 
\end{figure}

  A protocol for the BSMs matching consists of the following steps:
\begin{enumerate}
 \item Identify all possible connections between the vertices. Assign to each edge a weight $\Delta$ given by the absolute value of the difference between the arrival times. This quantity identifies the amount of decoherence that has been experienced by the older quantum memory. This is a meaningful quantity because one of the two involved quantum memories is always fresh.  The resulting data structure is a weighted bipartite graph  $\mathcal{A}=\{(e_j, \Delta_j)\}$, i.e. a set of edges (vertex-pairs) and edge weights.   
 \item For each weighted bipartite graph $\mathcal{A}$ solve the \emph{maximum cardinality bipartite matching} problem.  We denote by $\mathcal{M}$ the subset of $\mathcal{A}$ containing  graphs with exactly $\ell$ edges, where $\ell$ is the highest matching cardinality obtained over all graphs. This subset contains all graphs which maximize the number of connections, and thus the repeater rate.
 \item Select one element $A_i\in\mathcal{M}$ by the following strategies:
 \begin{enumerate}
 \item[Strategy 0]: choose with equal probability an arbitrary $A_i$ from $\mathcal{M}$
 \item[Strategy 1]: minimize the sum of the weights $\sum_{j=1}^{l} \Delta_j$
 \item[Strategy 2]: maximize the sum of the weights $\sum_{j=1}^{l} \Delta_j$ 
\end{enumerate}
\end{enumerate}
The optimization involved in strategy 1 and 2 is known in literature as \emph{maximum weighted bipartite matching}\cite{west2001introduction} and the optimization algorithm has  complexity  $O(m^3 \log m)$ \cite{liu2012distributable} where $m$ is the number of quantum memories on one side. 
 
Strategy 1 connects pairs which arrived with a short time difference giving as a result the highest correlations which are possible to  produce at a certain time. The disadvantage is that older pairs remain in the memories and therefore are used at a later time, having experienced decoherence for a long time.  Strategy 2 resolves this mentioned  problem, connecting pairs with the largest time difference. This strategy removes from the quantum memories older pairs as soon as possible, leaving only quantum memories  which suffered decoherence for a short time. The disadvantage is that poor correlations are produced even when perfect correlations could be obtained. We have seen therefore, that both strategies have advantages and disadvantages. In sec.~\ref{sec:decoherence} we will discuss  which strategy minimizes the memory requirements for QKD. However, note that the repeater rate is independent of the matching strategies.

\section{Repeater rate as function of connection length\label{sec:reprate}}
\newcommand{\tc}{T_{\mathrm{C}}}

\newcommand{\cone}{1}
\newcommand{\ctwo}{2}
\newcommand{\cc}{C}
\newcommand{\prob}{\textrm{Prob}}

In this section we derive analytical formulas for the repeater rate for the {\crmp }  and \frmp. Let $\tc$ be the current time (measured in time-bins as defined in the previous section), then the repeater rate is the fraction of successful BSMs per memory per time-bin calculated over the whole running time of the quantum repeater protocol, i.e.

\begin{equation}
\label{eq:rrep}
 R(\tc):=\frac{1}{\tc}\sum_{t=1}^{\tc} \frac{<\ell>(t_{2})}{m},
\end{equation}

where $<\ell(t_{2})>$ is the average number of successful BSMs at time $t_2$. Many quantum repeater protocols are based on a probabilistic Bell-state measurement \cite{sangouard_quantum_2011, duan_long-distance_2001}. The BSM is probabilistic when implemented with linear optics \cite{calsamiglia2001maximum} or with detectors of finite efficiency \cite{sangouard_quantum_2011}. When a measurement fails, the involved quantum memories are supposed to be emptied and this attempt is marked as unsuccessful. Let $P_{BSM}$ be the success probability of the BSM. The probability that $\ell$ BSMs are successful is given by
\begin{align}
\label{eq:probsigma}
 \prob&[\Sigma=\ell](t_2):=\nonumber\\&\sum_{i=\ell}^{m}\binom{i}{\ell}\prob[\Lambda=i](t_2)P_{BSM}^{\ell} (1-P_{BSM})^{i-\ell},
\end{align}
where $\Sigma$ and $\Lambda$ are  random variables that can assume values 0,1, ..., $m$. The random variable $\Sigma$ denotes the number of successful BSM and $\Lambda$ the number of performed BSMs. The factor $P_{BSM}^{\ell} (1-P_{BSM})^{i-\ell}$ represents the probability that $\ell$ BSM are successful and $i-\ell$ are not successful. This event can happen in $\binom{i}{\ell}$ different ways.

The average number of successful BSM at time $t_2$ is given by
\begin{equation}
\label{eq:avglprob}
 \avg{\ell}(t_2):=\sum_{\ell=0}^{m}\ell~\prob[\Sigma=\ell](t_2)
\end{equation}

In the following we will focus on $\prob[\Lambda]$.

\newcommand{\av}{\bold{a}}
\newcommand{\bv}{\bold{b}}
\newcommand{\cv}{\bold{c}}

\newcommand{\hsetpar}[2]{\mathcal{H}^{#1}_{#2}}
\newcommand{\hlpar}[3]{\hsetpar{#1}{#2}(#3)}
\newcommand{\hl}{\hlpar{m}{w}{\ell}}
\newcommand{\hset}{\hsetpar{m}{w}}

We denote as $\cv=(\av, \bv)$ one possible configuration of the quantum memories in the repeater station. The vectors $\av$ and $\bv$ of length $m$ represent the status of the quantum memories on Alice's and Bob's side, respectively (see fig.~\ref{fig:MultiplexingScheme}). Each component takes the value $0$ if the corresponding quantum memory is empty and $1$ otherwise. We define as $\hl$ the set of all configurations leading to $\ell$ BSMs where $w$ is the maximal connection length. For example, let $\av=(0, 1, 0, 1,1,0)$ and $\bv=(1,1, 0,0,0,1)$ be the configurations of the quantum memories as seen in fig.~\ref{fig:scheme} at $t_2$, then $(\av,\bv)\in\hlpar{6}{1}{2}$ and $(\av,\bv)\in\hlpar{6}{5}{3}$ but $(\av,\bv)\not\in\hlpar{6}{1}{3}$ because when $w=1$ the maximal number of connections is $\ell=2$. Moreover, the set of all possible configurations is $\hset:=\cup_{\ell=0}^{m}\hl$ .

We model the whole process consisting of storage and measurement with two maps. The \emph{storage map} $\sigma_{\ell}: \hset(0)\rightarrow \hset(\ell)$ connects configurations at time $t_0$, i.e., before photons are received,  to configurations at time $t_1$, i.e., after photons are received and stored. Given $\cv\in\hset(0)$, the probability to have the configuration $\cv'\in\hset(\ell)$ is given by
\begin{align}
 \prob[\sigma_{\ell}(\cv)=\cv']&:=\prob[\cv'|\cv] \\
 &:=\prod_{i=1}^{m}\prob[c'_{i}|c_{i}]\\
 &:=\prod_{i=1}^{m}\prob[a'_{i}|a_{i}]\prob[b'_{i}|b_{i}] \label{eq:probccp}, 
\end{align}
with 
\begin{align}
 \prob[a'_{i}|a_{i}]&:=(1-p)(1-a'_{i})(1-a_{i}) \nonumber \\
 &+ p a'_{i}(1-a_{i}) + a'_{i}a_{i} \label{eq:probaapbbp},
\end{align}
where $p$ is the probability that a photon has not been absorbed by the quantum channel. The probability $\prob[b'_{i}|b_{i}]$ is defined analogously. Equation \eqref{eq:probccp} holds because the channels connecting each source to each quantum memory are independent. Equation \eqref{eq:probaapbbp} gives the conditional probability to have a final configuration $a'_{i}$ starting from an initial configuration $a_{i}$. The three addends on the right-hand side of eq.~\eqref{eq:probaapbbp} are mutually exclusive, i.e. given a certain configuration at most one is not zero.

The measurement map $\mu_{\ell}:\hset(\ell)\rightarrow\hset(0)$ relates configurations at time $t_1$ and $(t+1)_0$, i.e., after the quantum memories have been used for the BSMs. This map is deterministic, as the configuration $\cv'\in\hset(\ell)$ after the measurement is uniquely determined by the matching algorithm.

\begin{figure}
 \includegraphics[width=.33\textwidth, angle=-90]{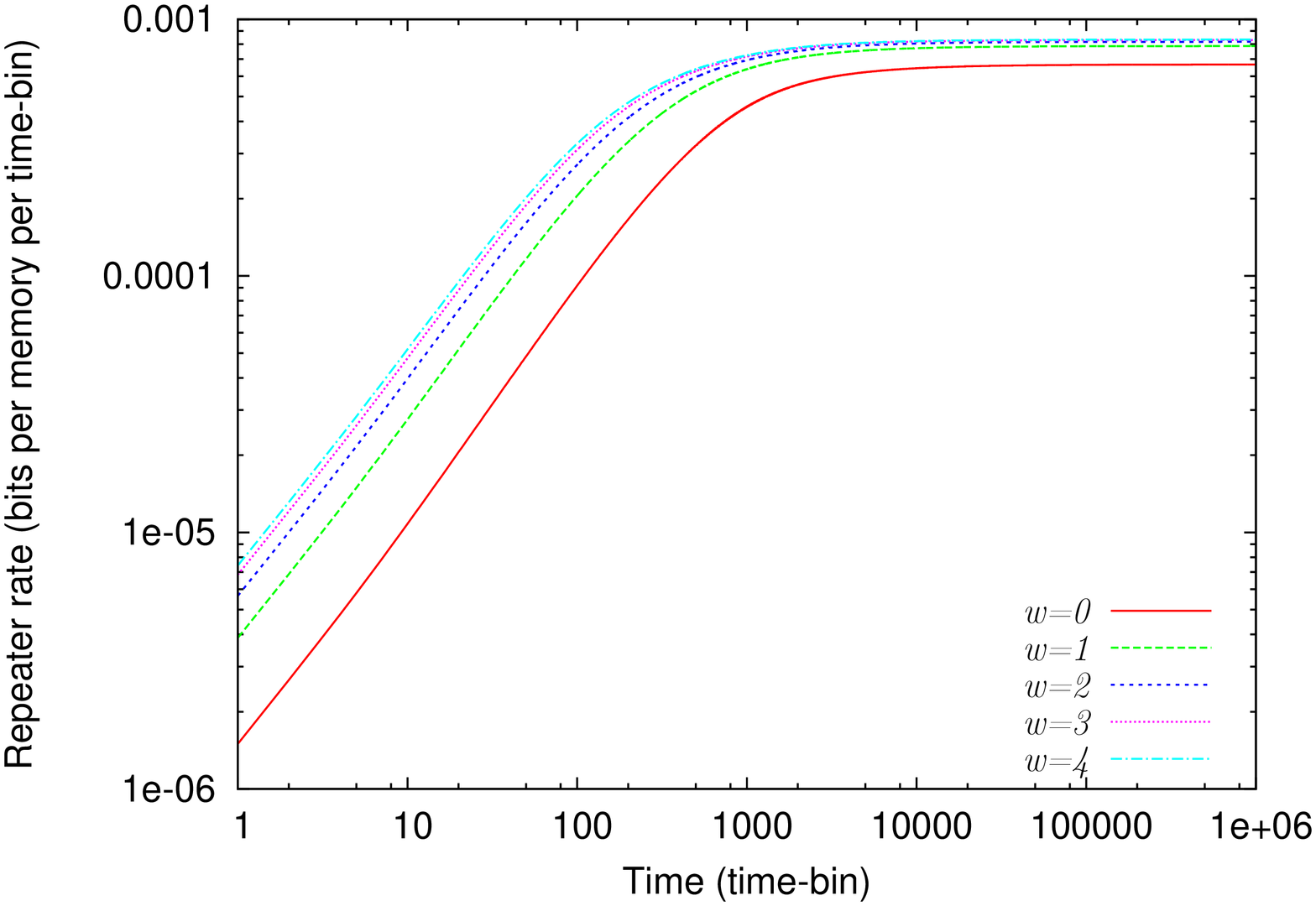}
\caption{(Color online) Repeater rate per memory as function of the time for $m=5$ and various maximal connection length $w$ (see eq.~\eqref{eq:rrep}). Parameters: $p=0.001, P_{BSM}=1$.}
\label{fig:multiplexingcomp} 
 \end{figure}

Coming back to  the probability $\prob[\Lambda=\ell](t_2)$ to have $\ell$ BSMs  we get
\begin{align}
 \prob[\Lambda=\ell](t_2)&:=\sum_{\cv'\in\hl}\prob[\cv'](t_1) \label{eq:probl}\\
 &=\sum_{\cv'\in\hl}\sum_{\cv\in\hset(0)}\prob[\sigma_{\ell}(\cv)=\cv']\prob[\cv](t_0),
\end{align}
which is the sum over all possible initial configurations $\cv$ and configurations $\cv'$ at time $t_2$  of the probability that $\cv$ leads to $\cv'$ weighted with the probability that the configuration $\cv$ was realized at time $t_0$. The probability $\prob[\cv](t_0)$ is given by 
\begin{widetext}
\begin{equation}
\label{eq:proct0}
 \prob[\cv](t_0)=\sum_{\cv'\in\hset(0)}\sum_{l=0}^{m}\prob[\mu_{\ell}\circ\sigma_{\ell}(\cv')=\cv]\prob[\cv']((t-1)_0),
\end{equation}
\end{widetext}
i.e., given a state $\cv'\in\hset(0)$ at time $(t-1)_0$, we calculate the probability that photon storage and measurement will lead to $\cv\in\hset(0)$ at time $t_0$. This last probability can be rewritten as
\begin{equation}
\label{eq:probt0expanded}
 \prob[\mu_{\ell}\circ\sigma_{\ell}(\cv)=\cv']=\sum_{\cv''\in\hset(\ell)}\delta_{\mu_{\ell}(\cv''),\cv'}\prob[\sigma_{\ell}(\cv)=\cv''],
\end{equation}
where $\delta_{\av, \bv}$ is the Kronecker delta such that $\delta_{\av, \bv}=1$ iff $\av=\bv$ and $\delta_{\av, \bv}=0$ otherwise.
Using the previous formulas and the initial condition $\prob[(\av, \bv)](0_0)=\delta_{\av,\bold{0}}\delta_{\bv,\bold{0}}$ we have now all elements for calculating the repeater rate in eq.~\eqref{eq:rrep}. In order to do that one inserts  eq.~\eqref{eq:probt0expanded} into eq.~\eqref{eq:proct0} which is then inserted in eq.~\eqref{eq:probl} which is finally used for calculating eq.~\eqref{eq:avglprob} through eq.~\eqref{eq:probsigma}. The calculation was performed in  C{}\verb!++!. The measurement map has been implemented using the library provided at Ref.~\cite{bworld}. The complexity of the calculation is proportional to the number of time steps because for calculating $<l>(t)$ it is sufficient to know quantities at time $t-1$. However, the set $\hl$ grows quite fast as function of $m$ and therefore reasonable time considerations restricted the calculation to a maximal of $m\leq7$. Regarding  fig.~\ref{fig:multiplexingcomp} the calculation ran for two days on a cluster of 10 nodes, cpu with four-cores and 
eight GB of RAM. 

In the following and for the rest of the paper we consider $p=0.001$ which represents the transmission probability of a single photon over an optical fiber of length $D=150$ km and for an absorption coefficient $\alpha=0.2$ dB/km. The relation between $p$ and the distance between Alice and the repeater station $D$ is $p=10^{-\frac{\alpha D}{10}}$ \cite{RevModPhys.74.145}.

As seen in fig.~\ref{fig:multiplexingcomp} the repeater rate increases as function of the time reaching a plateau at time $t\approx10^4$. We see that this behavior persists when changing the maximal connection length $w$. Analyzing the dependence on the maximal connection length, we observe in fig.~\ref{fig:multiplexingcomp} that the gap between $w=0$ and $w=1$ is almost the same than the gap between $w=0$ and $w=4$ which represents full-range multiplexing. This shows that in an experimental implementation in order to profit of multiplexing it is not necessary to have long-range connection. Moreover, for our set-up with a source at $1$ kHz, the loading time is $10$ s long. This result could give a hint that in more complex quantum repeater protocols with many repeater stations, using distillation and classical communication, the loading time could play a significant role in the total time of the  execution of the quantum repeater protocol. 

\begin{figure}
 \includegraphics[width=.33\textwidth, angle=-90]{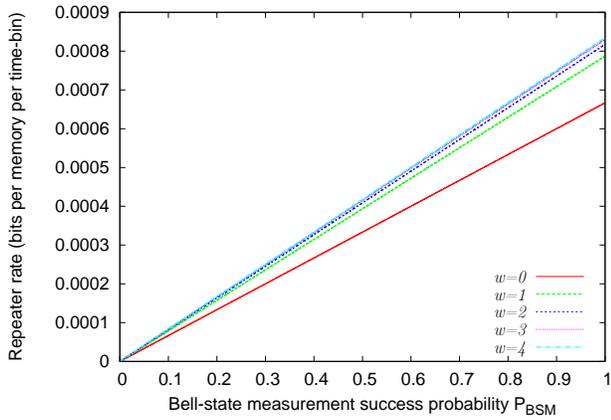}
\caption{(Color online) Repeater rate per memory per time-bin (eq.~\eqref{eq:rrep}, eq.~\eqref{eq:avglprob}, eq.~\eqref{eq:fx}) as a function of the BSM success probability. 
Parameters: $p=0.001, m=5$.}
\label{fig:ratepbsm} 
\end{figure}

Finally, we give an analytical formula for the rate when the plateau is reached. The time evolution of our system is specified in eq.~\eqref{eq:proct0}. For $t\rightarrow\infty$, the rate becomes time-independent:
\begin{equation}
\label{eq:proct0asympt}
 \prob[\cv]=\sum_{\cv'\in\hset(0)}\sum_{l=0}^{m}\prob[\mu_{\ell}\circ\sigma_{\ell}(\cv')=\cv]\prob[\cv'].
\end{equation}
Here $\prob[\cv]$ is the unknown to be determined.  We derive an analytical form of $\prob[\cv]$ in App.~\ref{app:fx}. Here, we use this formula (eq.~\eqref{eq:fx}) for calculating the asymptotic repeater rate as function of the BSM success probability. As shown in fig.~\ref{fig:ratepbsm}  the largest improvement is possible to have with full-range multiplexing, but already a similar improvement is reached with maximal connection length $w=1$, instead of $w=4$. Moreover, the linear behavior can be justified as follows. In the case of $p=0.001$ and for $w=1$ we obtain $\prob[\Lambda=1]=3.9\cdot10^{-3}$,  $\prob[\Lambda=2]=5.5\cdot10^{-6}$  and $\prob[\Lambda>2] \ll \prob[\Lambda=2]$.
 Therefore, eq.~\eqref{eq:avglprob} becomes $\avg{\ell}\approx P_{BSM} \prob[\Lambda=1]$ which is linear in $P_{BSM}$. For other values of the maximal connection length the order of magnitude of the probabilities is the same.

\section{Decoherence of the quantum memories\label{sec:decoherence}}
In case of multiplexing, even when the rate is maximized it is possible to connect pairs in different ways. In this section, we study the optimal matching algorithm for the BSM in relation to memory decoherence. For simplicity we will stick to deterministic entanglement swapping, i.e. $P_{BSM}=1$ in eq.~\eqref{eq:avglprob}.

Our figure of merit for optimizing the  matching algorithm will be the secret fraction, which in QKD characterizes the fraction of secret bits that can be extracted the from measured qubits \cite{Scarani:2009}. In the case of the BB84 protocol the secret fraction is given by \cite{Scarani:2009}
\begin{equation}
 r_{\infty}:=1-h(e_X)-h(e_Z), 
\end{equation}
where $e_X, e_Z$ are the quantum bit error rates (QBER) in base $X$ and $Z$ and $h(p):=-p\log_{2}p-(1-p)\log_{2}(1-p)$ is the binary Shannon entropy. For simplicity, we consider a symmetric error model such that $e_X=e_Z=:e$. The QBER resulting from measurements performed at time $t_2$ is 
\begin{equation}
 \overline{e}(t_2):=\sum_{\delta=0}^{t_2}\tilde{e}(\delta)\prob[\Delta=\delta](t_2),
\end{equation}
where $\prob[\Delta=\delta](t_2)$ is the fraction of measurements of quantum memories which have experienced decoherence for a time $\Delta=\delta$. This probability depends on the BSM strategy (see sec.~\ref{sec:protocol}). The QBER after these measurements is given by $\tilde{e}(\delta)$. This quantity depends on the decoherence mechanism of the quantum memories. In this paper we consider depolarization. Given $\rho_0$, the state of the quantum memory at time $t_0$, after depolarization it becomes
\begin{equation}
\label{eq:depmodel}
 \rho(t-t_0):=p(t-t_0)\rho_0+\frac{1-p(t-t_0)}{2}\one,
\end{equation}
where $p(t):=e^{-\frac{t}{\tau}}$ and $\tau$ is the decoherence time of the quantum memory. 

For the BB84 it holds\footnote{The reason is that entanglement swapping between two depolarized states with fidelities $F_0$ and $F_1$, respectively, will result in a depolarized state of fidelity $F_2=\frac{1}{3}(1-F_1-F_0+4F_0F_1)$. Inserting $F_1=1$ and $F_0=p$  and using the fact that $e=\frac{2}{3}(1-F_2)$ \cite{RevModPhys.74.145} the result follows.}
\begin{equation}
 \tilde{e}(\delta):=\frac{2}{3}(1-p(\delta)).
\end{equation}

The total QBER is calculated between all outcomes that Alice and Bob get from the beginning of the protocol until time $\tc$ which is equal to
\begin{equation}
\label{eq:etc}
 e(\tc):=\frac{\sum_{t=0}^{\tc}\avg{\ell}(t_2)\sum_{\delta=0}^{t_2}\tilde{e}(\delta)\prob[\Delta=\delta](t_2)}{\sum_{t=0}^{\tc}\avg{\ell}(t_2)}.
\end{equation}
Here, the denominator is the total number of BSMs from the beginning of the protocol until time $\tc$. The numerator is the average QBER for each time-bin weighted with the total number of successful measurements for each time-bin. The secret key rate is not zero whenever $e(\tc)\leq0.11$. This will be used to obtain a lower bound on the necessary coherence time $\tau$. 

We have calculated eq.~\eqref{eq:etc} using numerical simulations. It is also possible to proceed analytically as explained in sec.~\ref{sec:reprate}. However, the space of the configurations is so large that the analytical computation becomes unfeasible. We have performed  numerical simulations by repeating many times the protocol, and from the obtained connections we have calculated the averages. The number of used experiments is about $10^{9}$ which permits to have a variance of the mean smaller than $0.001$. The simulations were performed for the strategies $0,1$, and $2$, which were introduced in sec.~\ref{sec:protocol}.

\begin{figure}
 \includegraphics[width=.33\textwidth, angle=-90]{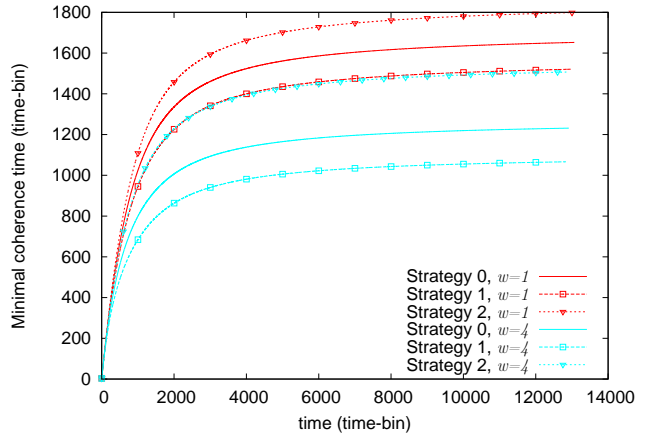}
\caption{(Color online) Minimal necessary coherence time $\tau$ (see eq.~\eqref{eq:depmodel}) as function of the time $t$ for different strategies. Strategy 0 (solid lines), strategy 1 (dashed lines with squares), strategy 2 (dotted lines with reversed triangles). Parameters: $p=0.001, m=5, w=1, 4.$ }
\label{fig:MinimalTau65} 
\end{figure}

As shown in fig.~\ref{fig:MinimalTau65} the minimal necessary coherence time is given by strategy 1. We see that strategy 0 is between strategy 1 and 2 and that the ordering between the strategy remains the same by changing the maximal connection length.

\begin{figure}
 \includegraphics[width=.33\textwidth, angle=-90]{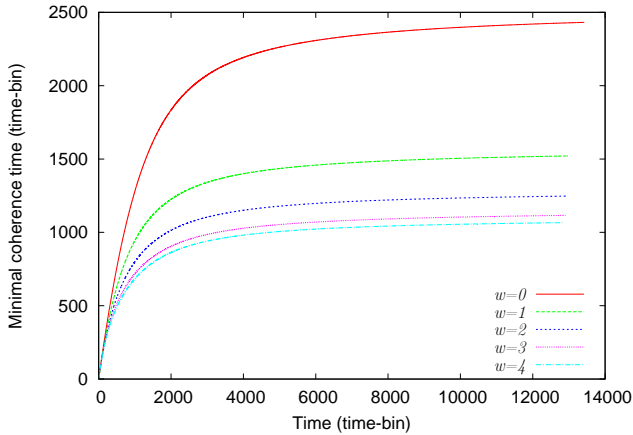}
\caption{(Color online) Minimal coherence time $\tau$ (see eq.~\eqref{eq:depmodel}) as function of the time $t$ for different maximal connection length.  Parameters: $p=0.001, m=5$. }
\label{fig:MinimalTauManyW} 
\end{figure}

We then studied how the minimal coherence time scales as function of $w$. As seen in fig.~\ref{fig:MinimalTauManyW} even a maximal connection length of $w=1$ has a significant impact on the minimal coherence time, compared to $w=0$. In particular we observe that for $t=12000$ the improvement from $w=0$ to $w=1$ is roughly $60\%$.

\section{Secret key rate\label{sec:scr}}
In this section we will sum up the results of the previous two sections and we will calculate the secret key rate. The secret key rate at time $\tc$ is calculated as the product of the repeater rate and the secret fraction \footnote{The sifting rate is not explicitly written because we assume that a biased choice of the bases is done \cite{Ardehali:fk}. Therefore the sifting rate in the asymptotic case is $1$.}, i.e. \cite{Scarani:2009, abruzzo2012quantum}
\begin{equation}
 K(\tc):=R(\tc)\cdot r_{\infty}(e(\tc)),
\end{equation}
where the repeater rate or raw key  rate $R(\tc)$ was defined in eq.~\eqref{eq:rrep} and the QBER was given in eq.~\eqref{eq:etc}. For our calculation we have used the minimal coherence time calculated in the previous section. In particular we have chosen
\begin{equation}
 \tau=5\tau_{\textrm{min}}(12800).
\end{equation}

\begin{figure}
 \includegraphics[width=.33\textwidth, angle=-90]{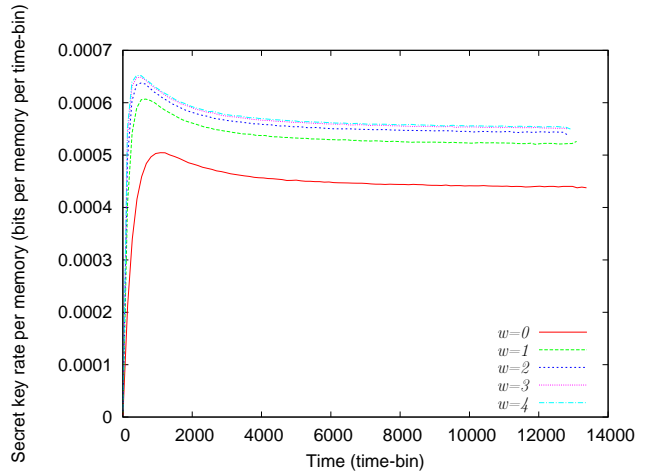}
\caption{(Color online) Secret key rate as function of the time $t$. for different maximal connection length.  Parameters: $p=0.001, m=5$. The coherence time is shown in tab.~\ref{tab:coherencetime}.}
\label{fig:ComparisonScr} 
\end{figure}

We use $t=12800$ because it is the highest achievable with our simulation (see fig.~\ref{fig:MinimalTauManyW}). In tab.~\ref{tab:coherencetime} we report the used coherence time for each value of $w$.  As shown in fig.~\ref{fig:ComparisonScr} the finite-range multiplexing with $w=1$ leads to a similar improvement as with $w=4$. Interestingly, we see that the secret key rate has a maximum for  $t<1000$ and then it slowly decreases. This fact can be explained by observing that there are two competing behaviors: the repeater rate  increases with the time and the secret fraction decreases with the time, as the  QBER increases with time, due to the fact that the probability that poor connections happen increases. An improvement may be to remove  very old pairs which are known  not to contribute to the final secret key. This method will certainly decrease the QBER at the expense to decrease also the repeater rate. We postpone to  future investigations new possible schemes which could permit to have a secret key 
rate which has a monotonic behavior. 

\begin{table}[h]
  \begin{tabular}{| c | c | c |}
  \hline
      $w$ & $\tau$ (time-bin)\\
      \hline\hline
      $0$ & $12133.02$\\
      $1$ & $8989.75$ \\
      $2$ & $7997.70$ \\
      $3$ & $7631.75$ \\
      $4$ & $7533.51$\\
        \hline
  \end{tabular}
    \caption{Value of the coherence time $\tau$ used for calculating the secret key rate. In order to obtain the value in seconds it is sufficient to divide by the frequency of the source. }
    \label{tab:coherencetime}
\end{table}

\section{Conclusions\label{sec:conclusions}}
Quantum repeaters offer the possibility to enlarge the distance where quantum key distribution becomes feasible. In this paper we have considered the scenario with one repeater station in the middle, containing several pairs of quantum memories. This is a generalization of measurement-device independent QKD with quantum memories and single-photon sources. We have considered multiplexing as a scheme for performing the Bell-state measurement. We introduced the concept of finite-range multiplexing which  originates from the experimental constraint that long range connections are demanding. We have characterized analytically the repeater rate for the case of probabilistic and deterministic Bell-state measurement. We  found that in a multiplexing protocol already short-range connections cover most of the improvement over a standard protocol. Decoherence of the quantum memories and different strategies for connecting the pairs were also studied. We found that it is always optimal to connect pairs with the shortest 
time difference in 
arrival time: this strategy minimizes the necessary coherence time required by the quantum memories in order to extract a secret key. Moreover, we have shown that also for the figure of merit ``minimal coherence time'' short-range multiplexing is almost as good as general multiplexing.  Finally, we have studied the secret key rate which characterizes the performance of quantum key distribution, finding results analogous to the previous sections. Future questions may include the case of Alice and Bob using weak coherent pulses. This can be done by following \cite{2013arXiv1306.3095A}. The analysis of finite-size effects for QKD can be performed by following \cite{PhysRevA.86.022332, PhysRevA.86.052305, 2013arXiv1307.1081C}.  The techniques derived in our paper may also be used  for addressing more complicated multiplexing protocols involving distillation and classical communication. Our work suggests  that in more complex protocols the loading time may play a significant role, thus reducing the  
repeater rate w.r.t. asymptotic formulas.

\section{Acknowledgment}
We would like to thank Sylvia Bratzik, Michael Epping, Ren\'e Schwonnek and Peter van Loock for valuable and enlightening discussions. We acknowledge financial support by the German Federal Ministry of Education and Research (BMBF, Project QuOReP).

\appendix

\section{Derivation of the repeater rate in the asymptotic case\label{app:fx}}
In order to determine the left-hand side of eq.~\eqref{eq:proct0asympt}, we define a function $\gamma:  0,1,...,N \rightarrow \hset(0)$ where $N=|\hset(0)|$ is the cardinality of $\hset(0)$. We rewrite eq.~\eqref{eq:proct0asympt} in the following way
\begin{equation}
\label{eq:proct0asymptf2}
 f(x)=\sum_{x'=0}^{N}q(x',x)f(x'),
\end{equation}
where $f(x):=\prob[\gamma(x)]$ and $q(x',x):=\sum_{l=0}^{m}\prob[\mu_{\ell}\circ\sigma_{\ell}(\gamma(x'))=\gamma(x)]$. The solution  is the following 
\begin{equation}
\label{eq:fx}
 f(x)=\frac{K_{N}(N,x)}{\sum_{x'=0}^{N}K_{N}(N,x')},
\end{equation}
with
\begin{widetext}
\begin{align}
\label{eq:kn}
 K_{N}(x',x)&:=\frac{K_{N-1}(N-1,x)}{1-K_{N-1}(N-1,N-1)}K_{N-1}(x', N-1) + K_{N-1}(x', x),\\
 K_{0}(x',x)&:=q(x',x).
\end{align}
\end{widetext}
In order to see that observe that eq.~\eqref{eq:proct0asymptf2} can be seen as a system of equations of the unknowns $\{f(0), f(1), ..., f(N)\}$, with the additional condition $\sum_{x=0}^{N}f(x)=1$ which comes from the fact that $f(x)$ is a probability and we sum over the whole space. Therefore we have
\begin{align}
 f(0) &= q(0,0)f(0)+\sum_{x'=1}^{N}q(x',0)f(x')\\
 &\Rightarrow f(x) = \sum_{x'=1}^{N}K_1(x',x)f(x'),
\end{align}
with $K_1(x',x)$ given in eq.~\eqref{eq:kn}. Repeating the procedure, the function $f(x)$ can be expressed as
\begin{equation}
 f(x):=K_{N}(N,x)f(N).
\end{equation}
The form in eq.~\eqref{eq:fx} is obtained by using the additional constraint $\sum_{x=0}^{N}f(x)=1$.

\end{document}